\documentclass[12pt,a4paper]{article}
\usepackage{latexsym}

\renewcommand{\footnotemark}{}

\begin{document}

\begin{center}
{\large\bf Physical interpretation of initial conditions \\[0.5ex]
for fractional differential equations \\[0.8ex]
with Riemann-Liouville fractional derivatives\footnote{Pre-print of the article published in
\emph{Rheologica Acta}, Online First, November 29, 2005). \\ The
original publication is available at www.springerlink.com \\
Original article DOI: 10.1007/s00397-005-0043-5.} }

\bigskip
\bigskip

{Nicole Heymans$^{(a)}$ and Igor Podlubny$^{(b)}$ \\[2ex]
$^{(a)}$Physique des Mat\'{e}riaux de Synth\`ese,\\
Universit\'e Libre de Bruxelles, \\
CP259, boulevard du Triomphe, 1050 Bruxelles,
Belgium \\ e-mail: nheymans@ulb.ac.be \\[1ex]
$^{(b)}$B.E.R.G. Faculty, Technical
University of Kosice, \\ B.~Nemcovej 3, 04200 Kosice, Slovak Republic \\
e-mail: igor.podlubny@tuke.sk}

\bigskip

September 1, 2005

\end{center}

\begin{abstract}
On a series of examples from the field of viscoelasticity
we demonstrate that it is possible to attribute physical meaning
to initial conditions expressed in terms of Riemann-Liouville
fractional derivatives, and that it is possible to obtain
initial values for such initial conditions by appropriate
measurements or observations.
\end{abstract}

\section[Introduction]{Introduction}

Many physical phenomena lead to their description in terms of non integer order differential equations. Formulations of non integer order derivatives, generally called fractional derivatives, fall into two main classes: Riem\-ann-Liouville derivatives and Gr\"{u}nwald-Letnikov derivatives, on one hand, defined as
(Podlubny 1999, Samko et al. 1993)
\begin{equation}
_{0}D_{t}^{\alpha} f(t) = \frac{1}{\Gamma (n-\alpha)}
\left(
\frac{d}{dt}
\right)^n
\!\!
\int\limits_{0}^{t}
\frac{f(\tau) \, d\tau}
     {(t-\tau)^{\alpha - n + 1}},
\end{equation}
or the Caputo derivative on the other, defined as (Caputo and Mainardi 1971)
\begin{equation}
_{0}^{C}D_{t}^{\alpha} f(t) = \frac{1}{\Gamma (n-\alpha)}
\!\!
\int\limits_{0}^{t}
\frac{ f^{(n)} (\tau) \, d\tau}
     {(t-\tau)^{\alpha - n + 1}},
\end{equation}
where $n-1 \leq \alpha < n$.

In this article we deal only with the Riemann-Liouville fractional derivatives.
Fractional differential equations in terms of the Riemann-Liouville derivatives  require initial conditions expressed in terms of initial values  of fractional derivatives of the unknown function (Podlubny 1999, Samko et al. 1993), like, for example, in the following initial value problem (where $n-1 < \alpha < n$):
\begin{equation}
_{0}D_{t}^{\alpha} f(t) + a f(t) = h(t);
\qquad \qquad
( t > 0 )
\end{equation}
\begin{equation}
\hspace*{2.5em}
\left[
\, _{0}D_{t}^{\alpha -k} f(t)
\right]_{t \rightarrow 0} = b_{k},
\qquad \qquad (k = 1, 2, \, \ldots \, , n).
\end{equation}

On the contrary, initial conditions for the Caputo derivatives are expressed in terms of initial values of integer order derivatives. It is known that for zero initial conditions the Riemann-Liouville, Gr\"{u}nwald-Letnikov and Caputo fractional derivatives coincide (Podlubny 1999). This allows a numerical solution of initial value problems for differential equations of non integer order independently of the chosen definition of the fractional derivative. For this reason, many authors either resort to Caputo derivatives, or use the Riemann-Liouville derivatives but avoid the problem of initial values of fractional derivatives by treating only the case of zero initial conditions.

It is frequently stated that the physical meaning of initial conditions expressed in terms of fractional derivatives is unclear or even non existent.
The old and ubiquitous requirement for physical interpretation of such initial conditions was most clearly formulated recently by
Diethelm et al. (2005):

\begin{quotation}
``A typical feature of differential equations (both classical and fractional) is the need to specify additional
conditions in order to produce a unique solution. For the case of Caputo FDEs, these additional conditions
are just the static initial conditions \ldots, which are akin to those of classical ODEs, and \emph{are therefore
familiar to us}. In contrast, for Riemann-Liouville FDEs, these additional conditions constitute certain
fractional derivatives (and/or integrals) of the unknown solution at the initial point $x = 0$ \ldots, which are
functions of $x$. These initial conditions \emph{are not physical}; furthermore, it is not clear how such quantities
are to be measured from experiment, say, so that they can be appropriately assigned in an analysis.''
\end{quotation}

(Emphasis is ours).
This quotation highlights the utmost importance of the interpretation of initial conditions in terms of fractional derivatives for further applications in various fields of science. The physical and geometric interpretations of operations of fractional integration and differentiation were suggested recently by Podlubny (2002). However, the problem of interpretation of initial conditions still remained open.

In this paper we shall show that initial conditions for fractional differential equations with Riemann-Liouville derivatives expressed in terms of fractional derivatives \emph{have physical meaning}, and that the corresponding quantities \emph{can be obtained from measurements}. We shall also demonstrate that in many instances of practical significance zero initial conditions, which are used so frequently in practice, appear in a natural way.

\section[Number of initial conditions, past history and memory]{Number of initial conditions, past history and memory}

When a physical process can be described in terms of a differential equation of integer order $n$, it is well known that $n$ conditions are required to solve the system. In this paper only initial conditions are considered. It is also known (Podlubny 1999, Samko et al. 1993) that fractional differential equations of order $\alpha$ require $\alpha*$ initial conditions, where $\alpha*$ is the lowest integer greater than $\alpha$. This means that if $\alpha<1$ as is the case in viscoelasticity when inertial effects are negligible, a single initial condition is sufficient. However, one of the reasons for the success encountered in describing viscoelasticity by means of differential equations of non integer order is their ability to describe real behaviour, including memory effects such as are observed in polymers, using only a restricted number of material parameters. Such memory effects may continue to affect the material response long after the cause has disappeared, as observed in stress relaxation after a non monotonous loading programme (Heymans and Kitagawa 2004). In such a case a single initial condition would appear insufficient to predict material response.

Here we shall consider only the response of a system starting at $t=0$ from a state of absolute rest. As a further simplification, we shall consider only response to ideal loading programs, such as step or impulse response. The effects of a finite loading time, of the details of the loading program, and
of past history will be accounted for separately in a sequel.

It has been shown (Beris and Edwards 1993) that thermodynamically valid constitutive equations for viscoelasticity are completely equivalent to analog models containing only elements (springs and dashpots) with positive coefficients. A suitable hierarchical arrangement of springs and dashpots gives rise to spring-pot behaviour (described below), either exactly at all timescales for an infinite tree (Heymans and Bauwens 1994), or in the long-term (or low-frequency) limit for an infinite ladder or infinite Sierpinski gasket (Heymans and Bauwens 1994, Schiessel and Blumen 1993, 1995). In the latter case short-term behaviour is similar to a Maxwell model with one element replaced by a spring-pot. Therefore the equivalence demonstrated by Beris and Edwards can be generalized to models including spring-pots (Heymans 1996), hence discussion here will be limited to such models.

\section[Spring-pot model]{Spring-pot model}

We shall start with a spring-pot alone, which is a linear viscoelastic element whose behaviour is intermediate between that of an elastic element (spring) and a viscous element (dashpot). The term ``spring-pot'' was introduced by Koeller (1984), although the concept of an element with intermediate properties had been introduced some time earlier.
The constitutive equation of a spring-pot is:

\begin{equation}
\sigma(t) = K\, _{0}D_{t}^{\alpha} \epsilon (t)
\qquad \mbox{or} \qquad
\epsilon (t) = \frac{1}{K} \,\, _{0}D_{t}^{-\alpha} \sigma(t)
\end{equation}

\noindent
where $\sigma$ is stress, $\epsilon$ is strain and $K$ is the model
constant. The spring-pot is the viscoelastic version of Westerlund's
``simplest model'' (Westerlund 2002). If $\alpha=0$ the element is linear
elastic (Hookean spring) whereas if $\alpha=1$ it is purely viscous
(Newtonian dashpot). Insight can be gained from the response of a
spring-pot in a few simple cases, using the general relationship (Podlubny
1999, Samko et al. 1993)
\begin{equation} _{0}D_{t}^{\alpha} (at^{p}) = a
\, \frac{\Gamma(1+p)} {\Gamma(1+p-\alpha)} t^{p-\alpha}.
\end{equation}

\subsection[Creep or general finite load]{Creep or general finite load}

In the case of creep, a stress step $\sigma_{0}$ is applied at initial time $t=0$.
The strain response is hence
$\epsilon(t) = (\sigma_{0} / K \Gamma(1+\alpha)) t^{\alpha}$.
The initial value of the strain vanishes, i.e. there is no instantaneous (elastic) strain, only an anelastic (retarded) response. However, the first ordinary derivative of strain is unbounded, so that a finite though undefined strain can be reached in an arbitrarily small time interval.

The change of $\epsilon(t)$ is described by the fractional differential equation
\begin{equation}\label{eq:SP-equation}
_{0}D_{t}^{\alpha} \epsilon(t) = \frac{\sigma_{0}}{K}
\end{equation}

In accordance with the theory of fractional differential equations
in terms of Riemann-Liouville derivatives,
an initial condition involving $_{0}D_{t}^{\alpha -1} \epsilon(t)$ is required. This condition can be found by taking the first-order integral of the constitutive equation as
\begin{displaymath}
\left[
\, _{0}D_{t}^{\alpha -1} \epsilon(t)
\right]_{t \rightarrow 0}
=
\left[
\, _{0}D_{t}^{-1}( \sigma_{0} / K)
\right]_{t \rightarrow 0}.
\end{displaymath}

In the case under consideration stress is finite at all times, hence \linebreak
$\left[
\, _{0}D_{t}^{-1} \sigma_{0} \right]_{t \rightarrow 0}=0$,
which leads to zero initial condition for $_{0}D_{t}^{\alpha -1} \epsilon(t)$, namely
\begin{equation}\label{eq:SP-Zero-IC}
\left[
\, _{0}D_{t}^{\alpha -1} \epsilon(t)
\right]_{t \rightarrow 0} = 0.
\end{equation}

The same considerations apply to a general finite load $\sigma(t)$. In the latter case the equation to be solved is
\begin{equation}\label{eq:SP-general-load}
_{0}D_{t}^{\alpha} \epsilon(t) = \frac{\sigma(t)}{K},
\end{equation}

\noindent
and the initial condition to be attached to this equation is the zero initial condition (\ref{eq:SP-Zero-IC}).

%%%%%%%%%%%%%%%%%%%%%%%%%%%%%%%%%%%%%%%%%%%%%%%%%%%%%%%
\subsection[Stress relaxation or general deformation]{Stress relaxation or general deformation}

The stress response to a strain step $\epsilon_{0}$ is
$\sigma(t) = (\epsilon_{0} \, K / \Gamma(1-\alpha)) t^{-\alpha}$.
The initial stress is unbounded reflecting the fact that a spring-pot (just like a dashpot) cannot respond immediately to a bounded stress: it has an infinite initial modulus or a vanishing initial compliance. However, relaxation to a finite though undefined stress occurs in an arbitrarily small time interval.

The change of  $\sigma(t)$ is described by the fractional differential equation
\begin{equation}
_{0}D_{t}^{-\alpha} \sigma(t) = K \epsilon_{0}.
\end{equation}

From the known value of $\epsilon_{0}$ we can obtain the initial value (as $t$ approaches zero) of
$_{0}D_{t}^{-\alpha}\sigma(t)$.
Clearly, if $_{0}D_{t}^{-\alpha}\sigma(t)$ is to be finite although it is defined over a vanishingly small time interval, $\sigma(t)$ must be unbounded. On the contrary, the initial value of $_{0}D_{t}^{-\alpha}\sigma(t)$ is well defined and finite, and that of $_{0}D_{t}^{-\alpha-1} \sigma(t)$ is zero. Thus, contrary to the idea expressed by some authors (e.g., Gl\"{o}ckle and   Nonnenmacher 1991), initial value problems expressed in terms of fractional integrals are not better posed than those expressed in terms of fractional derivatives.

If strain increases linearly with time, stress increases as $t^{1-\alpha}$. Stress is bounded, but the initial values of its integer order derivatives are unbounded. The known strain rate allows us to define the initial value of $_{0}D_{t}^{1-\alpha}\sigma(t)$. In fact, in this case, zero initial conditions are found both for $_{0}D_{t}^{-\alpha} \sigma(t)$ and $_{0}D_{t}^{-\alpha-1} \sigma(t)$.

For any general finite strain $\epsilon(t)$, following the same reasoning, again zero initial conditions are found.

In all three examples given here, initial conditions expressed in terms of fractional derivatives or integrals arise naturally when taking measurable quantities into account.

\subsection[Impulse response]{Impulse response}

The impulse response is seldom used in viscoelasticity except as a mathematical convenience, because it is even more problematic to apply a homogeneous impulse of stress or strain on a sample than it is to apply a step. However, we shall investigate the impulse response following the same reasoning as for the step response above.

Consider an impulse of stress defined as $B\delta(t)$ applied to the spring-pot at time $t=0$.
After that, the stress remains zero. The strain response is
$\epsilon(t) = (B / K \Gamma(\alpha)) t^{\alpha - 1}$.
The initial stress singularity gives rise to a lower-order strain singularity, since a spring-pot cannot deform immediately.

The strain $\epsilon(t)$ for $t>0$ is the solution to the fractional differential equation
\begin{equation}\label{eq:SP-impulse}
_{0}D_{t}^{\alpha}\epsilon(t) = 0.
\end{equation}

In accordance with the theory of fractional differential equations with Riemann-Liouville derivatives,
an initial condition involving
$\left[
\, _{0}D_{t}^{\alpha -1} \epsilon(t)
\right]_{t \rightarrow 0}$
is required.
This can be found through integration of the constitutive equation, as
\begin{displaymath}
\left[
\, _{0}D_{t}^{\alpha -1} \epsilon(t)
\right]_{t \rightarrow 0}
=
\left[
\, _{0}D_{t}^{-1} \sigma(t)/K
\right]_{t \rightarrow 0} = B/K,
\end{displaymath}
which gives the following initial condition to equation (\ref{eq:SP-impulse}):
\begin{equation}\label{eq:SP-impulse-IC}
\left[
\, _{0}D_{t}^{\alpha -1} \epsilon(t)
\right]_{t \rightarrow 0} = B/K.
\end{equation}

In this problem in terms of Riemann-Liouville derivatives $B$ is the initial impulse of stress $\sigma(t)$, $\epsilon(t)$ is the strain after application of this impulse, and the known impulse of stress yields a non-zero initial condition (\ref{eq:SP-impulse-IC}) involving a fractional derivative of strain. This fractional derivative is non zero, well defined, and bounded. Note that both strain and its integer-order derivatives are unbounded, and its first order integral is zero, so that a meaningful initial condition expressing the loading conditions cannot be obtained using integral-order derivatives.

The physically unrealistic stress response to a prescribed strain impulse will not be considered here. In fact, the analytical solution has a strong $t^{-(1+\alpha)}$ divergence, reflecting the fact that a strain impulse cannot be applied to a spring-pot.

\section[The key: look for inseparable twins]{The key: look for inseparable twins}

Now, after introducing the above simple example, let us formulate our general approach to interpretation of initial conditions involving the Riemann-Liouville fractional derivatives.

In a general case, when we consider some fractional differential equation for, say, $U(t)$, we have to consider also some function $V(t)$, for which some \emph{dual relation} exists between $U(t)$ and $V(t)$. For example, in viscoelasticity we have to consider the pair of stress $\sigma(t)$ and strain $\epsilon(t)$; in electrical circuits the pair of current $i(t)$
% (OR CHARGE $q(t)$)
and voltage $v(t)$; in heat conduction the pair of the temperature difference $T(t)$ and the heat flux $q(t)$; etc. Functions $U(t)$ and $V(t)$ are normally related by some basic physical law for the particular field of science. In each scientific field there are such pairs of functions like the aforementioned, which are as \emph{inseparable as Siamese twins}: the left-hand side of the initial condition involves one of them, whereas the evaluation of the right-hand side is related to the other.

This concept is not restricted to the spring-pot treated above, but is further applied in the subsequent sections to more elaborate models of viscoelastic behaviour. Indeed, a spring-pot is a particularly crude model, which has several unrealistic and unphysical characteristics. As pointed out above, it has a vanishing initial compliance or an infinite initial modulus. Viscoelastic solids, on the contrary, have a well-defined instantaneous modulus. (Note that an unbounded initial modulus is no more of a problem when describing a viscoelastic fluid than it is when describing Newtonian viscosity: if a step strain is applied to a dashpot, the initial stress is also unbounded). At long times there is no limit to anelastic strain of a spring-pot: creep continues indefinitely. Also, stress relaxes to vanishingly small values. The increase of stress in constant strain-rate conditions means that if attempting to describe a viscoelastic fluid, steady state flow is never attained. When describing a viscoelastic solid, again we find an unbounded modulus at $t=0$. In spite of these limitations, the spring-pot can give an approximate description of polymer viscoelasticity in the intermediate time range. Several slightly more elaborate models, which alleviate some oversimplifications of a single spring-pot, will be investigated below.

\section[The fractional order Voigt model]{The fractional order Voigt model}

The fractional Voigt model (a spring and a spring-pot in parallel) is nowadays generally understood as a long-term approximation to the fractional Zener model and it might seem irrelevant to express concern over initial value problems for the Voigt model. However, as the purpose of this note is mainly to examine how initial conditions endowed with physical meaning may be expressed in systems whose constitutive equations contain fractional derivatives, we shall continue to examine the fractional Voigt model.

The constitutive equation of this model is
\begin{equation}\label{eq:V-constitutive}
\sigma(t) = E \epsilon(t) + K \, _{0}D_{t}^{\alpha} \epsilon(t).
\end{equation}

The Voigt element or associations thereof are considered in viscoelasticity modelling to be appropriate to obtain the strain response to a prescribed stress program, so we shall investigate only such cases here.

Assume a stress impulse $B\delta(t)$ is applied to a Voigt element at time $t=0$. Then the fractional order equation we need to solve for $\epsilon(t)$ ($t>0$) is
\begin{equation}\label{eq:V-equation-stress-impulse}
E \epsilon(t) + K \, _{0}D_{t}^{\alpha} \epsilon(t) =0.
\end{equation}

In agreement with the theory of fractional differential equations in terms of Riemann-Liouville derivatives, we need an initial condition, which will involve the value of $_{0}D_{t}^{\alpha -1}\epsilon(t)$ for $t\rightarrow 0$. This condition can be obtained by integration of the constitutive equation as
\begin{equation}\label{eq:V-aux}
\left[
E \, _{0}D_{t}^{-1}\epsilon(t) + K \, _{0}D_{t}^{\alpha -1} \epsilon(t) = \, _{0}D_{t}^{-1}\sigma(t)
\right]_{t\rightarrow 0}.
\end{equation}

The limit of the right hand side is the magnitude $B$ of the stress impulse. On physical grounds, the spring-pot cannot deform instantaneously under a finite stress, and, as is the case for a spring-pot alone, any singularity of $\epsilon(t)$ must be weaker than that of the stress impulse, thus
\begin{displaymath}
\left[
\, _{0}D_{t}^{-1}\epsilon(t)
\right]_{t\rightarrow 0} = 0.
\end{displaymath}

This can also be found from examination of the behaviour of the left hand side of the relationship (\ref{eq:V-aux}): if $\left[
\, _{0}D_{t}^{-1}\epsilon(t)
\right]_{t\rightarrow 0}$ is non zero, then
$\left[
\, _{0}D_{t}^{\alpha-1}\epsilon(t)
\right]_{t\rightarrow 0}$ is unbounded and equation (\ref{eq:V-aux}) cannot be fulfilled. Hence the initial condition finally takes on the form of
\begin{equation}
\left[
K \, _{0}D_{t}^{\alpha-1}\epsilon(t)
\right]_{t\rightarrow 0} = B.
\end{equation}

This condition expresses the initial value of the fractional derivative of strain,  $_{0}D_{t}^{\alpha-1}\epsilon(t)$, in terms of the stress impulse. We see that the initial condition is obtained expressing a fractional derivative of the unknown strain in terms of a measurable, physically meaningful value of its ``inseparable Siamese twin''-- the stress. The obtained initial condition is, in fact, identical to the initial condition of the spring-pot alone. This reflects the known fact that the spring in the Voigt model only affects long-term behaviour.

Now let us consider creep, i.~e. the response to a stress step $\sigma_{0}$ applied at $t=0$. The equation to be solved for the strain $\epsilon(t)$ is
\begin{equation}\label{eq:V-creep-equation}
E \epsilon(t) + K \, _{0}D_{t}^{\alpha} \epsilon(t) = \sigma_{0},
\end{equation}

\noindent
and the initial condition for (\ref{eq:V-creep-equation}) can be found from
\begin{displaymath}
\left[
E \, _{0}D_{t}^{-1} \epsilon(t) + K \, _{0}D_{t}^{\alpha-1} \epsilon(t) = \, _{0}D_{t}^{-1} \sigma (t)
\right]_{t \rightarrow 0},
\end{displaymath}
where the limit of the right hand side is zero.
A bounded stress can produce only a bounded strain, so the limit of the first-order ordinary integral of strain in the left hand side is also zero. Thus the initial condition has the form:
\begin{equation}\label{eq:V-creep-IC}
\left[
\, _{0}D_{t}^{\alpha-1} \epsilon(t)
\right]_{t \rightarrow 0} = 0.
\end{equation}

Once more, knowledge of a measurable quantity ($\sigma_{0}$) leads to an initial condition expressed in terms of a fractional order derivative of the unknown ($\epsilon(t)$), its inseparable Siamese twin.

The case of a general finite stress program is similar to that of creep. The equation to be solved is now
\begin{equation}
E \epsilon(t) + K \, _{0}D_{t}^{\alpha} \epsilon(t) = \sigma(t),
\end{equation}

\noindent
and the initial condition is identical to the condition (\ref{eq:V-creep-IC}) obtained in creep.

\section[The fractional order Maxwell model]{The fractional order Maxwell model}

To keep to a simple model while describing realistic behaviour for a viscoelastic solid, a spring expressing instantaneous elasticity must be associated in series with the spring-pot. This eliminates the unbounded initial stress in description of relaxation. The Maxwell element or associations thereof are considered in viscoelasticity modelling to be appropriate to obtain the stress response to a prescribed strain program, so we shall investigate only such cases here.

The constitutive equation of the Maxwell model is
\begin{displaymath}
\epsilon (t) = \frac{1}{E} \, \sigma(t) + \frac{1}{K} \, (\, _{0}D_{t}^{-\alpha} \sigma(t)),
\end{displaymath}
or
\begin{equation}\label{eq:M-constitutive}
\frac{1}{E} \, _{0}D_{t}^{\alpha}\sigma(t)
+ \frac{1}{K} \,  \sigma(t)
= \, _{0}D_{t}^{\alpha} \epsilon(t).
\end{equation}

In stress relaxation, a step strain $\epsilon_{0}$ is applied at $t=0$. Then the equation to be solved is
\begin{equation}\label{eq:M-relax-equation}
\frac{1}{E} \, _{0}D_{t}^{\alpha}\sigma(t)
+ \frac{1}{K} \,  \sigma(t)
=
\frac{\epsilon_{0} t^{-\alpha}}
     {\Gamma (1-\alpha)}.
\end{equation}

An initial condition is required, involving
the value of $\left[ \, _{0}D_{t}^{\alpha-1}\sigma(t) \right]_{t \rightarrow 0}$.
Integrating the constitutive equation (\ref{eq:M-constitutive}) and considering the limit as $t$ approaches zero, we have
\begin{equation}\label{eq:M-initial-condition}
\left[
\frac{1}{E} \, _{0}D_{t}^{\alpha-1}\sigma(t)
+ \frac{1}{K} \,  _{0}D_{t}^{-1}\sigma(t)
= \, _{0}D_{t}^{\alpha-1} \epsilon(t)
\right]_{t \rightarrow 0}.
\end{equation}

Since strain remains bounded during loading, and $\alpha<1$, the right hand side inside brackets is bounded and vanishes when $t \rightarrow 0$. Since the left hand side is a linear combination of positive functions with positive coefficients, it can only vanish if each term vanishes. This means that stress remains bounded during loading, and hence that we obtain the following initial condition:
\begin{equation}
\left[
\, _{0}D_{t}^{\alpha -1} \sigma(t)
\right]_{t \rightarrow 0} = 0.
\end{equation}

Here again we observe that
the initial condition on the unknown stress arises naturally from its Siamese twin, the known strain.

Now we shall consider the strain impulse response. A strain impulse of magnitude $A\delta(t)$ is applied at time t=0. Thereafter, the equation to be solved is
\begin{equation}\label{eq:M-impulse-equation}
\frac{1}{E} \, \sigma(t)
+ \frac{1}{K} \,  _{0}D_{t}^{-\alpha}\sigma(t)
= 0.
\end{equation}

The required initial condition is obtained as above by
integrating the constitutive equation (\ref{eq:M-constitutive}) and considering the limit as $t$ approaches zero:

\begin{equation}\label{eq:M-IC-Impulse}
\left[\frac{1}{E} \, _{0}D_{t}^{-1}\sigma(t)
+ \frac{1}{K} \, _{0}D_{t}^{-\alpha-1}\sigma(t)
= \, _{0}D_{t}^{-1} \epsilon(t)
\right]_{t \rightarrow 0}.
\end{equation}

The limit of the right hand side of (\ref{eq:M-IC-Impulse}) is A.
Hence the limit of the left hand side must also be bounded.
This means that the limit of the first-order integral of stress must be bounded,
and the $\alpha+1$ integral must vanish, and the initial condition is
finally
\begin{equation}
\left[\frac{1}{E} \, _{0}D_{t}^{ -1}\sigma(t) \right]_{t \rightarrow 0}
= A.
\end{equation}

The singularity in the stress response to a strain impulse is now of the same order as that of the strain impulse itself: adding a spring in series with the spring-pot has weakened the singularity.

The strain response to a stress impulse is identical to that of a spring-pot alone since the impulse response of the spring vanishes.

\section[The fractional order Zener model]{The fractional order Zener model}

Among the fractional order models of viscoelasticity considered in this article, the most general is the Zener model. Its constitutive equation is
\begin{equation}
\sigma(t) + \nu \; _{0}D_{t}^{\alpha} \sigma(t)
=
\lambda \, \epsilon(t) + \mu \; _{0}D_{t}^{\alpha} \epsilon(t),
\end{equation}
where $\lambda = E_{\infty}$ is the long-term modulus,
$\mu=K (E_0-E_\infty)/{E_0}$,
$\nu=\mu / E_{0}$
and $E_0$ is the instantaneous modulus.

Let us first investigate the response to a stress impulse $B\delta(t)$ applied to the Zener element at time $t=0$.
Then the fractional differential equation we need to solve for $\epsilon(t)$ ($t>0$) is:
\begin{equation}\label{eq:Z-gen}
\lambda \epsilon(t) + \mu \, _{0}D_{t}^{\alpha} \epsilon(t) = 0.
\end{equation}

In accordance with the theory of fractional differential equations, we need an initial condition involving the initial value of $_{0}D_{t}^{\alpha-1} \epsilon(t)$.
Integration of the constitutive equation gives:
\begin{equation}\label{eq:Z-auxiliary}
_{0}D_{t}^{-1}\sigma(t) + \nu \, _{0}D_{t}^{\alpha-1} \sigma(t)
=
\lambda \, _{0}D_{t}^{-1}\epsilon(t) + \mu \, _{0}D_{t}^{\alpha-1} \epsilon(t).
\end{equation}

The initial condition can be found by considering equation (\ref{eq:Z-auxiliary}) as $t \rightarrow 0$:
\begin{equation}\label{eq:Z-IC}
\left[
\, _{0}D_{t}^{-1}\sigma(t) + \nu \, _{0}D_{t}^{\alpha-1} \sigma(t)
=
\lambda \, _{0}D_{t}^{-1}\epsilon(t) + \mu \, _{0}D_{t}^{\alpha-1} \epsilon(t)
\right]_{t \rightarrow 0}.
\end{equation}

Using considerations similar to those in case of the Voigt model under stress impulse, we obtain the initial condition in the form:
\begin{equation}
\left[
\mu \, _{0}D_{t}^{\alpha-1} \epsilon(t)
\right]_{t \leftrightarrow 0}
= B
\end{equation}

As in case of the Voigt model, this condition gives the initial value of the fractional derivative of unknown strain,  $_{0}D_{t}^{\alpha-1}\epsilon(t)$, in terms of its ``inseparable twin'' -- the stress.

The right and left hand sides of equations (\ref{eq:Z-gen}) and (\ref{eq:Z-IC}) are formally identical, hence following the same reasoning as above the response to a strain impulse $A\delta(t)$ applied to the Zener element at time $t=0$ will be the solution to the equation
\begin{equation}
\sigma(t) + \nu \; _{0}D_{t}^{\alpha} \sigma(t) = 0
\end{equation}
with the initial condition
\begin{equation}
\left[
\nu \, _{0}D_{t}^{\alpha-1} \sigma(t)
\right]_{t \leftrightarrow 0}
= A
\end{equation}

This formal equivalence between response to a stress or strain impulse reflects the well known fact that the Zener model is the simplest model capable of describing response to a stress or strain program equally well.

In case of creep, i.e. a step-stress $\sigma(t)=\sigma_{0}$ for $\sigma>0$, we have the equation
for $\epsilon(t)$:
\begin{equation}\label{eq:Z-creep_equaton}
\lambda \epsilon(t) + \mu \, _{0}D_{t}^{\alpha} \epsilon(t)
=
\sigma_{0} + \nu \, \sigma_{0}\, \frac{t^{-\alpha}}{\Gamma(1-\alpha)}.
\end{equation}

The initial condition can also be found by considering equation (\ref{eq:Z-auxiliary}) as $t \rightarrow 0$.

Following a similar reasoning to that given above for the Maxwell model in stress relaxation, we find a zero initial condition to accompany equation (\ref{eq:Z-creep_equaton}):
\begin{equation}
\left[ \,
_{0}D_{t}^{\alpha-1} \epsilon(t)
\right]_{t \rightarrow 0} = 0.
\end{equation}

This initial condition in terms of fractional derivative of $\epsilon(t)$ appeared from consideration of its ``inseparable twin'' $\sigma(t)$.

Similarly, for stress relaxation, $\epsilon(t)=\epsilon_{0}$, we obtain the equation for $\sigma(t)$:
\begin{equation}
\sigma(t) + \nu \, _{0}D_{t}^{\alpha} \sigma(t)
=
\lambda \epsilon_{0} + \mu \, \epsilon_{0}\, \frac{t^{-\alpha}}{\Gamma(1-\alpha)}.
\end{equation}

The initial condition for the unknown stress $\sigma(t)$,
\begin{equation}
\left[ \,
_{0}D_{t}^{\alpha-1} \sigma(t)
\right]_{t \rightarrow 0} = 0,
\end{equation}

\noindent
appears naturally from consideration of the initial value of strain.

Let us now consider the case of general load $\sigma(t)=\sigma_{\ast} (t)$.
The equation to be solved for $\epsilon(t)$ is
\begin{equation}
\lambda \epsilon(t) + \mu \, _{0}D_{t}^{\alpha} \epsilon(t)
=
\sigma_{\ast}(t) + \nu \, _{0}D_{t}^{\alpha} \sigma_{\ast}(t)
\end{equation}

The corresponding initial condition can be obtained using the following procedure. Consider some small $t=a$. Starting at $t=0$, stress $\sigma(t)$ must be recorded until $t=a$, and based on the recorded values the left hand side of the relationship (\ref{eq:Z-auxiliary}) must be evaluated. The obtained quantity provides an approximation of the initial value for the expression in the right hand side of (\ref{eq:Z-auxiliary}).

In some cases it is possible to find the limit of such approximation as $a \rightarrow 0$. For example, for a physically realisable continuous load $\sigma_{\ast}(t)$ we obtain a zero initial condition in the form:
\begin{equation}
\left[
_{0}D_{t}^{\alpha -1} \epsilon(t)
\right]_{t \rightarrow 0} = 0.
\end{equation}

It is worth mentioning that this procedure amounts, in fact, to the same as measuring the initial value of, for example, the first derivative in the case of classical differential equations of integer order. From the examples given above, it can be seen that for any physically realistic model, zero initial conditions will be found for a continuous loading program or even in the case of a step discontinuity. Non-zero conditions will only be found in the case of an impulse.

\section[Conclusion]{Conclusion}

In this note we demonstrated on a series of examples that it is possible to attribute physical meaning to initial conditions expressed in terms of Riemann-Liouville fractional derivatives.

To summarize, expressing initial conditions in terms
of fractional derivatives of a function $U(t)$ is not a problem, because it does not require
a direct experimental evaluation of these fractional derivatives.
Instead, one should consider its ``inseparable twin'' $V(t)$
related to $U(t)$ via a basic physical law, and measure (or
consider) its initial values.

It is worth noting that the only case where non zero initial conditions appeared in our considerations, is the case of impulse response. In other cases (including the Zener model under physically realisable load program), the initial conditions are zero, and in such cases the use of the Riemann-Liouville derivatives, the Gr\"{u}nwald-Letnikov derivatives, and the Caputo derivatives is equivalent.

\end{document}